\documentstyle[prd,aps,epsf,12pt]{revtex}
\begin{document} 

\tighten
\draft
\preprint{DAMTP-97-134} 

\title{Thermalisation of Quantum States} 
 
\author{
Dorje C. Brody$^{*}$ 
and 
Lane P. Hughston$^{\dagger}$ 
} 
\address{$*$DAMTP, Silver Street, Cambridge CB3 9EW U.K.} 
\address{$\dagger$Merrill Lynch International, 
25 Ropemaker Street, London EC2Y 9LY U.K. \\ 
and King's College London, The Strand, London WC2R 2LS, U.K.} 

\date{\today} 

\maketitle 

\begin{abstract} 
An exact stochastic model for the thermalisation of quantum states 
is proposed. The model has various physically appealing properties. 
The dynamics are characterised by an underlying Schr\"odinger 
evolution, together with a nonlinear term driving the system 
towards an asymptotic equilibrium state and a stochastic term 
reflecting fluctuations. There are two free parameters, one of 
which can be identified with the heat bath temperature, while the 
other determines the characteristic time scale for thermalisation. 
Exact expressions are derived for the evolutionary dynamics of the 
system energy, the system entropy, and the associated density 
operator. 
\end{abstract} 
 
\pacs{PACS Numbers : 05.70.Ln, 02.50.Ey, 02.40.Vh, 05.30.Ch} 


\section{Introduction} 

The construction of a microscopic description for dynamical systems 
out of equilibrium is a topic of considerable interest in 
statistical mechanics. While many studies have been pursued in 
this direction \cite{reichl}, the 
subject remains refractory owing to its nonlinear character. In 
particular, a realistic framework must be simple 
enough to be tractable, yet sufficiently sophisticated to capture 
some of the physical essentials. Our goal here is to propose in 
this spirit an elementary but compelling stochastic model that 
characterises the thermalisation process for a quantum system in an 
arbitrary given initial state.\par 

The approach here, though carried out largely within the framework of 
standard quantum theory, involves a reexamination of the traditional 
hypotheses forming the basis of quantum statistical mechanics. In 
particular, we shall take the view that a quantum statistical 
ensemble, or `state', is correctly described not simply by a density 
matrix $\rho^{\alpha}_{\beta}$, but rather by a probability 
distribution on the space of 
pure quantum mechanical states. For measurements associated with 
ordinary quantum mechanical observables this description reduces 
to the usual one, for which the unconditional expectation 
of a linear observable is given by the familiar trace formula 
involving $\rho^{\alpha}_{\beta}$. The advantage of the present 
formulation, however, 
is that we can give meaning to the concept of ensembles 
and other mixed states in nonlinear quantum mechanics 
\cite{kibble,weinberg}. 
Furthermore, the fact that the space of pure states has a symplectic 
structure allows us to introduce 
the concept of a quantum phase space ${\sl\Gamma}$, and hence the 
use of probability distributions on ${\sl\Gamma}$ as ensembles. In 
particular, for thermal equilibrium in certain situations we propose 
the distribution given in (\ref{eq:cano0}) below, which we call the 
canonical ${\sl\Gamma}$-ensemble. The corresponding density matrix 
differs in some respects from the conventional density matrix of 
quantum statistical mechanics, and thus the applicability of the 
theory may in principle be testable. \par 

Our first step is to derive the Fokker-Planck equation for the 
time-dependent probability density function associated with a 
general Brownian motion with drift on the space of pure quantum 
states. Since ${\sl\Gamma}$ is compact, we can apply a 
theorem of Zeeman to show that there is a special class of drift 
terms for which, given any initial condition, the resulting 
asymptotic stationary solution is the canonical 
${\sl\Gamma}$-ensemble described above. The nonequilibrium dynamics 
we consider here is governed by a nonlinear Schr\"odinger evolution 
coupled to a thermal noise term, described by the stochastic 
differential equation (\ref{eq:drift}). This choice is the simplest 
possible that leads to thermal equilibrium for arbitrary initial 
states. \par 

In what follows we use the dynamical law (\ref{eq:drift}) to derive 
a formula for the total system energy (\ref{eq:U}), leading to a 
formula for the heat capacity (\ref{eq:C}). 
These ideas are applied, by way of illustration, to the case of a 
spin one-half particle in a heat bath. Then we derive an expression 
for the dynamics of the system entropy, given in 
(\ref{eq:dSdt}), from which we are able to deduce a fundamental 
thermodynamic inequality (\ref{eq:dSdt2}). Finally we study the 
evolution of the density matrix associated with the 
underlying process. We show that the time development of the density 
matrix involves a higher moment of the projection operator onto 
pure states. \par 

\section{Quantum Hamiltonian dynamics} 

We begin by considering the quantum mechanical phase space $CP^{n}$ 
\cite{kibble}. This is the space of pure quantum states, 
given by rays through the origin in the associated Hilbert space. 
The quantum phase space can be viewed as a $2n$-dimensional real 
manifold ${\sl\Gamma}$, endowed with the Fubini-Study metric 
$g_{ab}$. The 
reason that ${\sl\Gamma}$ plays the role of a quantum phase space is 
that Hamilton's equations on ${\sl\Gamma}$ can be lifted to the 
Hilbert space to give the Schr\"odinger dynamics. This 
follows from the fact that ${\sl\Gamma}$ has a natural symplectic 
structure, given by a nondegenerate antisymmetric tensor 
$\Omega_{ab}$, compatible with the metric $g_{ab}$ in the 
sense that $\nabla_{a}\Omega_{bc}=0$ and $\Omega_{ab}\Omega^{bc} = 
-\delta^{c}_{a}$, where $\Omega^{ab} = g^{ac}g^{bd}\Omega_{cd}$ and 
$\nabla_{a}$ denotes the standard covariant derivative, satisfying 
$\nabla_{a}g_{bc}=0$. Then the Schr\"odinger dynamics take the 
Hamiltonian form 
\begin{equation} 
dx^{a}\ =\ 2\Omega^{ab}\nabla_{b}Hdt\ , 
\end{equation} 
where 
$dx^{a}/dt$ is tangent to the quantum phase space trajectory. 
The Hamiltonian function $H(x)$ on ${\sl\Gamma}$ is given by 
the expectation of the energy operator at each point 
$x\in{\sl\Gamma}$. More specifically, the expectation of the 
Hermitian 
operator $H^{\alpha}_{\beta}$ with respect to the Hilbert space 
vector $\psi^{\alpha}$ $(\alpha, \beta = 0, 1, \cdots, n)$ is given 
by $H(x) = H^{\alpha}_{\beta}{\bar \psi}_{\alpha}\psi^{\beta}/
{\bar \psi}_{\gamma}\psi^{\gamma}$. Since each point $x \in 
{\sl\Gamma}$ corresponds to an equivalence class $\{ \lambda 
\psi^{\alpha}, \lambda\in{\bf C}-0\}$ for some Hilbert space vector 
$\psi^{\alpha}$, it follows that $H(x)$ is defined 
globally on ${\sl\Gamma}$. Conversely, one can show that such 
functions correspond to global solutions of the equation 
\begin{equation} 
\nabla^{2}H\ =\ (n+1)({\bar H}-H)\ , 
\end{equation} 
where $\nabla^{2}$ is the Laplace-Beltrami operator on ${\sl\Gamma}$. 
Here ${\bar H}=H^{\alpha}_{\alpha}/(n+1)$, $H^{\alpha}_{\alpha}$ 
being the trace of the Hamiltonian. Thus ${\bar H}$ is the uniform 
average of the energy eigenvalues, corresponding to the thermal 
equilibrium energy of the system at high temperature. \par 

\section{Stochastic dynamics} 

Now we generalise the Schr\"odinger dynamics by consideration of a 
diffusion process $x_{t}$ taking values in ${\sl\Gamma}$. For this 
purpose we make use of standard techniques of stochastic differential 
geometry \cite{i-w,hughston}. We shall examine the case of a Brownian 
motion with drift on ${\sl\Gamma}$, determined by the covariant 
stochastic differential equation  
\begin{equation} 
dx^{a}\ =\ \mu^{a}dt + \kappa \sigma^{a}_{i}dW^{i}_{t}\ , 
\label{eq:stoc} 
\end{equation} 
where $\kappa$ is a constant, $\mu^{a}$ is a vector field, and 
the vectors $\sigma^{a}_{i}$ $(i=1,2,\cdots, 2n)$ constitute an 
orthonormal basis in the tangent space of ${\sl\Gamma}$ such that 
$g^{ab} = 
\sigma^{a}_{i}\sigma^{b}_{j}\delta^{ij}$ and $\sigma^{a}_{i} 
\sigma^{b}_{j}g_{ab}=\delta_{ij}$. Here $dx^{a}$ is the covariant 
Ito differential \cite{hughston}, and the standard $2n$-dimensional 
Wiener process $W^{i}_{t}$ satisfies $dW^{i}_{t}dW^{j}_{t} = 
\delta^{ij}dt$. In local coordinates $x^{\bf a}$ $({\bf a} = 
1,2,\cdots,2n)$, the Ito differential is given by 
\begin{equation} 
dx^{a}\ =\ \delta^{a}_{\bf a}(dx^{\bf 
a}+\frac{1}{2}\kappa^{2}\Gamma^{\bf a}_{\bf bc}g^{\bf bc}dt)\ . 
\label{eq:covito} 
\end{equation} 
Here $\delta^{a}_{\bf a}$ is a coordinate basis for the given patch 
of ${\sl\Gamma}$, and $\delta^{\bf a}_{a}$ the dual basis, such that 
for the covariant derivative $\nabla_{a}\xi^{b}$ of a vector field 
$\xi^{a}$ with components $\xi^{\bf a} = \delta^{\bf a}_{a}\xi^{a}$ 
we can write 
\begin{equation} 
\delta^{b}_{\bf b}\delta^{\bf a}_{a}(\nabla_{b}\xi^{a})\ =\ 
\frac{\partial\xi^{\bf a}}{\partial x^{\bf b}} 
+ \Gamma^{\bf a}_{\bf bc}
\xi^{\bf c}\ . 
\end{equation} 
The Ito differential (\ref{eq:covito}) is constructed in such a way 
that the stochastic differential equation (\ref{eq:stoc}) is fully 
tensorial. \par 

Suppose $\phi(x)$ is a smooth function on 
${\sl\Gamma}$, and define the process $\phi_{t}=\phi(x_{t})$. 
It follows from Ito's lemma that $d\phi_{t} = 
\nabla_{a}\phi dx^{a}+\frac{1}{2}\nabla_{a}\nabla_{b}\phi 
dx^{a}dx^{b}$. Then the relation $dx^{a}dx^{b} = 
\kappa^{2}g^{ab}dt$ implies 
\begin{equation} 
d\phi_{t}\ =\ (\mu^{a}\nabla_{a}\phi + \frac{1}{2} \kappa^{2}
\nabla^{2}\phi)dt + \kappa \nabla_{a}\phi\sigma^{a}_{i}dW^{i}_{t}\ , 
\label{eq:ito} 
\end{equation} 
where $\nabla^{2}\phi=g^{ab}\nabla_{a}\nabla_{b}\phi$. Since the 
process $x_{t}$ is Markovian, there is a normalised density function 
$\rho(x,t)$ for the state at time $t$, characterised by a partial 
differential equation subject to specified initial conditions. The 
expectation of the process $\phi_{t}$ is thus 
\begin{equation}   
E[\phi_{t}]\ =\ \int_{\sl\Gamma}\phi(x)\rho(x,t) dV\ , 
\label{eq:exp} 
\end{equation} 
where $dV$ denotes the volume element on ${\sl\Gamma}$. On the other 
hand, it follows from Ito's lemma (\ref{eq:ito}) that 
\begin{eqnarray} 
\phi_{t}\ =\ \phi_{0} + 
\int_{0}^{t}\left( \mu^{a}\nabla_{a}\phi + 
\frac{1}{2}\kappa^{2}\nabla^{2}\phi\right) ds + 
\kappa\int_{0}^{t} \sigma^{a}_{i}\nabla_{a}\phi\, dW^{i}_{s}\ . 
\end{eqnarray} 
Here the integrands are valued at time $s$ at the point 
$x_{s}\in{\sl\Gamma}$. Since the second integral above has vanishing 
expectation, we obtain 
\begin{equation} 
E[\phi_{t}]\ =\ \phi_{0} + E\left[ 
\int_{0}^{t}(\mu^{a}\nabla_{a}\phi + \frac{1}{2}\kappa^{2}
\nabla^{2}\phi)ds\right] \ . \label{eq:Ephi} 
\end{equation} 
Hence differentiating (\ref{eq:exp}) and (\ref{eq:Ephi}) with 
respect to $t$ and equating the results, we have 
\begin{eqnarray} 
\int_{\sl\Gamma} {\dot \rho}(x,t)\phi(x)dV = 
\int_{\sl\Gamma}(\mu^{a}\nabla_{a}\phi+\frac{\kappa^{2}}{2}
\nabla^{2}\phi)\rho(x,t)dV, \label{eq:fp0} 
\end{eqnarray} 
where ${\dot \rho}=\partial\rho/\partial t$. Integrating by 
parts, and using the fact that the resulting relation 
must hold for all $\phi(x)$, we find that $\rho(x,t)$ satisfies 
\begin{equation} 
\frac{\partial}{\partial t}\rho(x,t)\ =\ -\nabla_{a}(\mu^{a}\rho) 
+ \frac{1}{2}\kappa^{2}\nabla^{2}\rho\ . \label{eq:f-p} 
\end{equation} 
This is the covariant Fokker-Planck equation for the density 
function on ${\sl\Gamma}$, corresponding to the stochastic 
process (\ref{eq:stoc}). The solution of (\ref{eq:f-p}) thus 
characterises the distribution of the diffusion (\ref{eq:stoc}) at 
time $t$, given an initial distribution $\rho(x,0)$. In the case of 
a singular initial distribution, e.g., an initially pure state, we 
interpret $\rho(x,t)$ as a generalised function, i.e., it has to 
satisfy (\ref{eq:fp0}) for all smooth $\phi(x)$. \par 

The foregoing results are valid on arbitrary compact Riemannian 
manifolds. We are concerned, however, with the case where 
${\sl\Gamma}$ is the quantum phase space $CP^{n}$, endowed with the 
Fubini-Study metric. Since ${\sl\Gamma}$ is compact, equation 
(\ref{eq:f-p}) may admit nontrivial asymptotic ($t\rightarrow\infty$) 
stationary solutions. We shall examine a simple situation in which 
this is the case, namely, when the drift is given by a 
gradient flow $\mu^{a}=-\frac{1}{2}\kappa^{2}\beta\nabla^{a}H$ 
generated by $H(x)$, where $\beta$ is a parameter. Then we can use 
a theorem of Zeeman \cite{zeeman} to show that there is a unique 
stationary solution for (\ref{eq:f-p}) of the form 
\begin{equation} 
\rho(x)\ =\ \frac{\exp(-\beta H(x))}{Z(\beta)}\ , \label{eq:cano0} 
\end{equation} 
where $Z(\beta)=\int_{\sl\Gamma}\exp(-\beta H(x))dV$. 
It follows that the probability density function $p(E)$ for the 
distribution of the energy function $H(x)$ is given by the formula 
$p(E) = Z^{-1}(\beta)\Omega(E)\exp(-\beta E)$, 
where 
\begin{equation} 
\Omega(E)\ =\ \int_{\sl\Gamma}\delta(H(x)-E)dV 
\end{equation} 
is the density 
of states in ${\sl\Gamma}$ for which $E\leq H(x)< E+dE$, and 
$Z(\beta)$ is the partition function. 
Another way of characterising the distribution (\ref{eq:cano0}) is 
that it maximises the entropy 
$S=-\int_{\sl\Gamma}\rho(x)\ln\rho(x)dV$ for 
a given value of system energy $U=\int_{\sl\Gamma}\rho(x)H(x)dV$. 
The theorem of Zeeman implies in the present context that, 
(\ref{eq:cano0}) is the asymptotic distribution of the process 
$x_{t}$ under essentially arbitrary initial conditions, and that 
$p(E)$ is the asymptotic energy distribution. \par 

We have so far considered a process of the form (\ref{eq:stoc}) 
with the drift $\mu^{a}=-\frac{1}{2}
\kappa^{2}\beta\nabla^{a}H$. This process, as such, does not yet 
make reference to the Schr\"odinger dynamics. If we view the 
drift term as a nonlinear correction to the 
underlying quantum evolution, we can represent the complete 
dynamics according to the prescription 
\begin{equation} 
dx^{a}\ =\ \left( 2\Omega^{ab}\nabla_{b}H - 
\frac{1}{2}\kappa^{2}\beta\nabla^{a}H\right) dt + 
\kappa \sigma^{a}_{i}dW^{i}_{t}\ , \label{eq:drift} 
\end{equation} 
which in the limit $\kappa\rightarrow0$ reduces to the ordinary 
Schr\"odinger dynamics. Due to the antisymmetry of $\Omega_{ab}$, 
inclusion of the symplectic term does not affect the resulting 
asymptotic state (\ref{eq:cano0}), since $\rho(x)$ is a function 
of $H(x)$. Hence, the analysis above shows that, given an arbitrary 
initial state (pure or general) on the quantum phase space, the 
dynamical law (\ref{eq:drift}) necessarily takes that state into 
the thermal equilibrium state (\ref{eq:cano0}). In particular, the 
process (\ref{eq:drift}) is ergodic on the energy surface, and 
asymptotically leads to a uniform distribution on each such surface. 
We note that (\ref{eq:drift}) involves two parameters, $\beta$ and 
$\kappa$. The stationary solution for $\rho(x,t)$ depends only on 
$\beta$, which we identify as the inverse of the heat bath 
temperature. The parameter $\kappa$, which has the units $s^{-1/2}$, 
and may depend on $\beta$, determines the thermalisation time 
scale. \par 

\section{Hamiltonian process} 

Our next step is to study the stochastic process $H_{t}=H(x_{t})$ 
associated with the Hamiltonian function. From Ito's lemma 
(\ref{eq:ito}) for the process (\ref{eq:drift}) we find  
\begin{equation} 
dH_{t}\ =\ \frac{1}{2}\kappa^{2}(\nabla^{2}H_{t} - \beta V_{t})dt 
+ \kappa\nabla_{a}H\sigma^{a}_{i}dW^{i}_{t}\ . \label{eq:hamil} 
\end{equation} 
Here $V_{t}=V(x_{t})$, where $V(x) = g^{ab}\nabla_{a}H\nabla_{b}H$ 
is the squared energy uncertainty at the point $x\in{\sl\Gamma}$, 
conditional on the pure state to which $x$ corresponds. Integrating 
(\ref{eq:hamil}) and taking its expectation, we have  
\begin{equation} 
E[H_{t}] = H_{0}+\frac{1}{2}\kappa^{2}\int_{0}^{t}
E\left[ \nabla^{2}H_{s} - \beta V_{s} \right] ds. 
\label{eq:hamil2} 
\end{equation} 
Let $U_{t}$ denote $E[H_{t}]$, the unconditional energy expectation, 
which can be interpreted as the total system energy at time $t$. 
Then by differentiating (\ref{eq:hamil2}) we obtain 
\begin{equation} 
\frac{\partial U_{t}}{\partial t}\ =\ \frac{1}{2}\kappa^{2} 
\left( (n+1)({\bar H} - U_{t}) - \beta E[V_{t}] \right)\ , 
\label{eq:dUdt} 
\end{equation} 
by use of the Laplace-Beltrami equation for $H_{t}$. This relation 
can be integrated to yield the solution for the time development 
of the system energy: 
\begin{eqnarray} 
U_{t}\ &=&\ {\bar H} + (H_{0} - {\bar H})e^{-\frac{1}{2}
\kappa^{2}(n+1)t} \nonumber \\ 
& &\ -\frac{1}{2}\kappa^{2}\beta \int_{0}^{t} \int_{\sl\Gamma}
e^{\frac{1}{2}\kappa^{2}(n+1)(s-t)} V(x) \rho(x,s) 
dV ds . \label{eq:U}
\end{eqnarray} 
In the limit $t\rightarrow\infty$ the only contribution is given by 
the trace term ${\bar H}$ and the integral of $V(x)$, and the 
resulting energy approaches the internal energy of the system in 
thermal equilibrium. In particular, in the high temperature limit 
$\beta \rightarrow 0$ the contribution from $V(x)$ vanishes, and 
we recover the uniform average of the energy eigenvalues. In the low 
temperature limit $\beta\rightarrow\infty$, the gradient term in the 
drift of (\ref{eq:drift}) dominates, and the system is forced 
to fall to the ground state. It is interesting to observe, as a 
consequence of (\ref{eq:dUdt}), that once thermal equilibrium is 
reached, we have the identity 
\begin{equation} 
kT^{2}C\ =\ {\rm Var}[H] + (n+1)kT(U-{\bar H})\ ,\label{eq:C}  
\end{equation} 
for the heat capacity $C=\partial U/\partial T$, where $\beta=1/kT$ 
and ${\rm Var}[H] = (\Delta H)^{2}_{\rho}$ is the unconditional 
energy variance. This follows from the conditional 
variance formula 
\begin{equation} 
{\rm Var}[H]\ =\ E\left[ {\rm Var}_{x}[H]\right] 
+ {\rm Var}\left[ E_{x}[H]\right]\ , 
\end{equation} 
where $E_{x}[H]$ and 
${\rm Var}_{x}[H]$ are, respectively, the conditional expectation 
and the conditional variance of the energy, when the system 
is in the pure state $x$. This relation expresses the total energy 
uncertainty by the sum of terms corresponding to quantum and 
thermal uncertainties. \par 

As an illustration, consider the case of a spin one-half particle 
in heat bath. The state space is $CP^{1}$, which we view as a 
2-sphere, the north and the south poles corresponding to the upper 
and lower energy eigenstates, with energies $+h$ and $-h$, where $h$ 
is the magnetic moment of the particle times the external field 
strength. The symplectic flow gives rise to latitudinal circular 
orbits on the sphere, while the gradient flow is in the direction 
along the great circles passing through the two eigenstates, pushing 
the state towards the south pole. The equilibrium energy latitude is 
obtained by balancing the gradient flow and the Brownian fluctuations. 
If we let $\theta$ denote the angular coordinate for the state 
as measured from the north pole, then the squared energy uncertainty, 
given that the system is in a pure state at latitude $\theta$, is 
${\rm Var}_{\theta}[H] = h^{2}\sin^{2}\theta$. The conditional energy 
expectation is $E_{\theta}[H] = h\cos\theta$. Since ${\bar H}$ 
vanishes, the only contribution to the energy in (\ref{eq:U}) is 
from the volume integral, which gives $U=\beta^{-1}-h\coth(\beta h)$, 
in agreement with the result of a direct calculation of 
$E[E_{\theta}[H]]$. At infinite temperature the system 
energy corresponds to that of the equator, whereas for finite 
temperature the system energy corresponds to that of a latitude in 
the southern hemisphere. At zero temperature, the system collapses 
to the ground state. \par 

\section{Entropy production} 

Next we consider the dynamics of the total entropy of the system, 
given by $S_{t}=-\int_{\sl\Gamma}\rho(x,t)\ln\rho(x,t)dV$. 
Without loss of generality we can write 
\begin{equation} 
\rho(x,t)\ =\ \frac{\exp\left( -\beta( H(x)+\eta(x,t))\right)} 
{\int_{\sl\Gamma}\exp\left( -\beta( H(x)+\eta(x,t))\right)dV}\ , 
\end{equation} 
for a general density function, where $\eta(x,t)$ determines the 
departure from thermal equilibrium. Substituting this expression 
into the formula for $S_{t}$, and making use of 
the Fokker-Planck equation (\ref{eq:f-p}) with drift as in 
(\ref{eq:drift}), we find 
\begin{equation} 
\frac{\partial S_{t}}{\partial t}\ =\ \frac{1}{2}\kappa^{2} 
\beta^{2}\int_{\sl\Gamma}(\nabla^{a}\eta\nabla_{a}H + 
\nabla^{a}\eta\nabla_{a}\eta)\rho dV\ . \label{eq:Sdot} 
\end{equation} 
On the other hand, for the total energy $U_{t}$ we have 
\begin{equation} 
\frac{\partial U_{t}}{\partial t}\ =\ 
\frac{1}{2}\kappa^{2}\beta\int_{\sl\Gamma} 
(\nabla^{a}\eta\nabla_{a}H)\rho dV\ . 
\end{equation} 
Therefore, we obtain 
\begin{equation} 
\frac{\partial S_{t}}{\partial t}\ =\ \beta 
\frac{\partial U_{t}}{\partial t} + \frac{1}{2}\kappa^{2}\beta^{2} 
\int_{\sl\Gamma}(\nabla^{a}\eta\nabla_{a}\eta)\rho dV\ . 
\label{eq:dSdt} 
\end{equation} 
This is the general formula for the entropy production associated 
with the process (\ref{eq:drift}), which shows that 
\begin{equation} 
\frac{\partial S_{t}}{\partial t}\ \geq\ \beta 
\frac{\partial U_{t}}{\partial t} \label{eq:dSdt2} 
\end{equation} 
in all circumstances. In particular, if $U_{t}$ increases as a 
consequence of heating, then the entropy of the system also 
necessarily increases. \par 

\section{Evolution of the density matrix}  

Finally we derive the dynamics of the density matrix 
$\rho^{\alpha}_{\beta}(t)$ associated with the process 
(\ref{eq:drift}). Let us revert to homogeneous coordinates 
for the state space $CP^{n}$, and write $\Pi^{\alpha}_{\beta}(x) = 
\psi^{\alpha}(x){\bar \psi}_{\beta}(x)/\psi^{\gamma}(x)
{\bar \psi}_{\gamma}(x)$ for the projection operator onto the pure 
state $x\in{\sl\Gamma}$ represented by the Hilbert space vector 
$\psi^{\alpha}(x)$. From equations (\ref{eq:exp}) and (\ref{eq:fp0}) 
it follows that for a general (possibly nonlinear) observable 
$\phi(x)$ on ${\sl\Gamma}$ we have 
\begin{eqnarray} 
\frac{\partial}{\partial t}E[\phi(x_{t})]\ &=&\ \int_{\sl\Gamma} 
\left( 2\Omega^{ab}\nabla_{a}\phi\nabla_{b}H + 
\frac{1}{2}\kappa^{2}\nabla^{2}\phi \right. \nonumber \\ 
& &\ \left. - 
\frac{1}{2}\kappa^{2}\beta g^{ab}\nabla_{a}\phi\nabla_{b}H 
\right) \rho(x,t)dV. \label{eq:phidot} 
\end{eqnarray} 
However, if $\phi(x)$ is given, more specifically, by the conditional 
expectation of an ordinary linear quantum mechanical observable 
$\phi^{\alpha}_{\beta}$, then $\phi(x)=\phi_{\alpha}^{\beta}
\Pi^{\alpha}_{\beta}(x)$. Thus for a linear observable we have 
$E[\phi(x_{t})] = \phi_{\alpha}^{\beta}\rho^{\alpha}_{\beta}(t)$, 
where 
\begin{equation} 
\rho^{\alpha}_{\beta}(t)\ =\ \int_{\sl\Gamma} \Pi^{\alpha}_{\beta}(x) 
\rho(x,t) dV \label{eq:deno} 
\end{equation} 
is the time dependent density matrix associated with the state 
$\rho(x,t)$. To obtain the equation of motion for 
$\rho^{\alpha}_{\beta}(t)$ we need to evaluate the three terms 
in the integrand on the right of (\ref{eq:phidot}). 
These are given as follows. For the commutator term we have 
\begin{equation} 
\Omega^{ab}\nabla_{a}\phi\nabla_{b}H\ =\ \frac{1}{2}i( 
\phi_{\gamma}^{\beta}H^{\gamma}_{\alpha} - H_{\gamma}^{\beta}
\phi^{\gamma}_{\alpha})\ , 
\end{equation} 
for the term involving the Laplace-Beltrami operator we find 
\begin{equation} 
\nabla^{2}\phi\ =\ 
(\phi_{\gamma}^{\gamma}\delta^{\beta}_{\alpha} - (n+1)
\phi_{\alpha}^{\beta})\Pi^{\alpha}_{\beta}\ , 
\end{equation} 
and for the anticommutator term we obtain 
\begin{equation} 
g^{ab}\nabla_{a}\phi\nabla_{b}H\ =\ 
\frac{1}{2}(\phi_{\gamma}^{\beta}H^{\gamma}_{\alpha} + 
H_{\gamma}^{\beta}\phi^{\gamma}_{\alpha})\Pi^{\alpha}_{\beta} - 
(\phi^{\beta}_{\alpha}H^{\delta}_{\gamma}) 
\Pi^{\alpha}_{\beta}\Pi^{\gamma}_{\delta}\ . 
\end{equation} 
Inserting these expressions into (\ref{eq:phidot}), and noting 
that the result must hold for arbitrary 
$\phi^{\beta}_{\alpha}$, we deduce that 
\begin{eqnarray} 
\frac{\partial}{\partial t}\rho^{\alpha}_{\beta}\ &=&\ 
i\left[ H^{\alpha}_{\gamma}\rho^{\gamma}_{\beta} - 
\rho^{\alpha}_{\gamma}H^{\gamma}_{\beta}\right] - 
\frac{1}{4}\kappa^{2}\beta \left( H^{\alpha}_{\gamma}
\rho^{\gamma}_{\beta} + 
\rho^{\alpha}_{\gamma}H^{\gamma}_{\beta}\right) \nonumber \\ 
& &\ + \frac{1}{2}\kappa^{2}\left( \delta^{\alpha}_{\beta} 
- (n+1)\rho^{\alpha}_{\beta} + \beta H_{\gamma}^{\delta}
\rho^{\alpha\gamma}_{\beta\delta}\right). \label{eq:liouville}  
\end{eqnarray} 
Here the density matrix $\rho^{\alpha}_{\beta}(t)$ is given by 
(\ref{eq:deno}), the first moment of the projection operator 
$\Pi^{\alpha}_{\beta}(x)$, whereas 
$\rho^{\beta\delta}_{\alpha\gamma}(t)$ 
is the second moment of $\Pi^{\alpha}_{\beta}(x)$, given by 
\begin{equation} 
\rho_{\beta\delta}^{\alpha\gamma}(t)\ =\ 
\int_{\sl\Gamma} \Pi^{\alpha}_{\beta}(x)\Pi^{\gamma}_{\delta}(x)
\rho(x,t) dV\ . 
\end{equation} 
Equation (\ref{eq:liouville}) is the dynamical law for the 
density operator, which takes an arbitrary initial state 
into an equilibrium state, with heat bath temperature $T$. 
The first term on the right of (\ref{eq:liouville}) leads to 
the Liouville equation of linear quantum dynamics, while the 
general nonequilibrium process has a richer structure. The 
emergence of the second moment term in (\ref{eq:liouville}) can 
also be interpreted by analogy with the renormalisation group 
equations. \par 

\section{Discussion} 

The model we have described here is surprisingly tractable, and 
has many attractive features. Experimental support could be 
pursued in two stages. First, we point out that it is not 
difficult, in the case of the canonical ${\sl\Gamma}$-ensemble, 
to derive explicit formulae for the 
partition function $Z(\beta)$, the state density $\Omega(E)$, 
the density matrix $\rho^{\alpha}_{\beta}$, and the second 
moment $\rho^{\alpha\gamma}_{\beta\delta}$. If these results 
turn out to give a better account of equilibrium phenomena than 
the conventional approach in some context, then the next step 
would be to look 
for effects resulting from the higher moment term in the 
nonequilibrium dynamical law (\ref{eq:liouville}). \par 

DCB is grateful to PPARC for financial support. \par 

$*$ Electronic address: d.brody@damtp.cam.ac.uk \par 
$\dagger$ Electronic address: lane@ml.com\par

\end{document}